\documentclass[twocolumn,showpacs,preprintnumbers,amsmath,amssymb]{revtex4}

\usepackage{graphicx}
\usepackage{dcolumn}
\usepackage{bm}

\begin{document}


\title{Non-invasive detection of charge-rearrangement in a quantum dot in high magnetic fields}

\author{C. Fricke$^{1}$} \author{M. C. Rogge$^{1}$} \author{B. Harke$^{1,2}$} \author{M. Reinwald$^{3}$} \author{W. Wegscheider$^{3}$} \author{F. Hohls$^{1,4}$} \author{R. J. Haug$^{1}$}
 \affiliation{$^{1}$Institut f\"ur Festk\"orperphysik, Universit\"at Hannover,
  Appelstra\ss{}e 2, D-30167 Hannover, Germany \\ $^{2}$ Max-Planck-Institut f\"ur biophysikalische Chemie, G\"ottingen, D-37077 \\ $^{3}$ Angewandte und
Experimentelle Physik, Universit\"at Regensburg, D-93040
Regensburg, Germany\\
$^{4}$Cavendish Laboratory, University of Cambridge,
  Madingley Road, Cambridge CB20HE, U, Great Britain}%

\date{\today}

\begin{abstract}

We demonstrate electron redistribution caused by magnetic field on a single quantum dot measured by means of a quantum point contact as non-invasive detector. Our device which is fabricated by local anodic oxidation allows to control independently the quantum point contact and all tunnelling barriers of the quantum dot. Thus we are able to measure both the change of the quantum dot charge and also changes of the electron configuration at constant number of electrons on the quantum dot. We use these features to exploit the quantum dot in a high magnetic field where transport through the quantum dot displays the effects of Landau shells and spin blockade. We confirm the internal rearrangement of electrons as function of the magnetic field for a fixed number of electrons on the quantum dot. 

\end{abstract}

\maketitle


Soon after their first realization quantum dots were seen as a model system for the study of the inner structure of a few electron system \cite{Kouwenhoven-97}. They were often called artificial atoms in relation to their natural counterpart. Electronic transport through the dots allowed for quite a sophisticated investigation of e.g. the shell structure of few electron dots \cite{Kouwenhoven-Tarucha}. But changes of the internal structure i.e. a charge redistribution without changes of the electron number are rather difficult to access.

The most recent interest in quantum dots stems from the goal of realizing quantum bits (qubits) in a semiconductor structure. Readout schemes for these qubits require non-invasive methods of charge and spin detection \cite{field-93}. Here quantum point contacts (QPC) can be used to detect individual tunneling events of electrons out of the quantum dot (QD) or between the dots of a double dot system \cite{elzerman_nature-04, elzerman_prb-03, schleser-04, petta-04, dicarlo-04}.

Here we will demonstrate that the QPC can also be used to detect changes of the electron configuration of a QD without changing the number of electrons. Our interest focuses on a QD in a high magnetic field. Transport measurements through a QD in this regime display an interesting behavior of the chemical potential of the QD and the tunneling amplitudes which were interpreted by McEuen et al in a framework of Landau shells \cite{McEuen-91}. Here we will give direct evidence of the redistribution of charge on the QD within the Coulomb blockade regime which confirms the interpretation of McEuen et al.


Our device is based on a GaAs/AlGaAs heterostructure containing a two-dimensional electron system (2DES) 34 nm below the surface. The electron density is $n = 4.59 \cdot 10^{15} \hspace{1.2mm}\mathrm{m}^{-2}$, the mobility is $\mu = 64.3 \hspace{1.2mm} \mathrm{m^2/V s}$. We use an atomic force microscope (AFM) to define the QD and the QPC structure by local anodic oxidation (LAO) \cite{held-98, ullik-00, pepper_afm-04}. In this way the 2DES below the oxidized surface is depleted and insulating areas can be written.
\begin{figure}[b]
\includegraphics[scale=0.8]{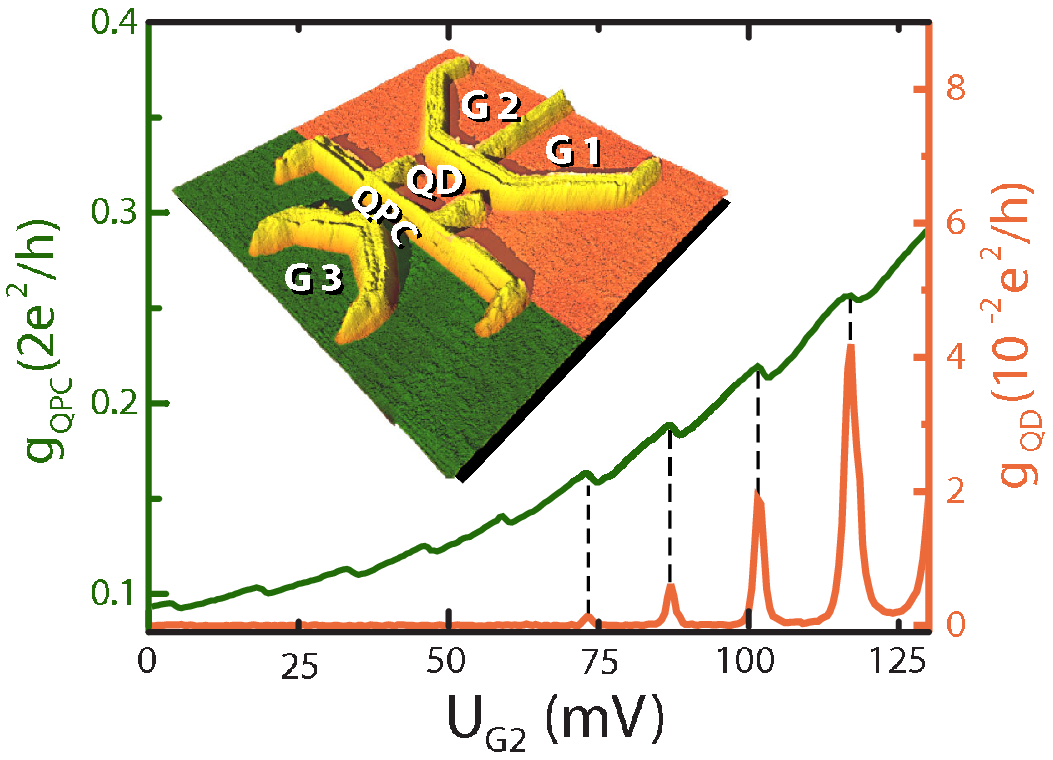}
\caption{\label{fig:qpcprinzip} Operating principle of the device containing a QD and a QPC. Conductivity of QD (lower line, right axis) and QPC (upper line, left axis) are shown as a function of gate voltage applied to G2. The inset shows a three-dimensional AFM image of the device.}
\end{figure}

An AFM image of our device is presented in the inset of Fig. \ref{fig:qpcprinzip}. The bright walls depict the insulating lines written by the AFM. The QPC (left area) is separated from the QD structure (right area) by an insulating line. The QPC can be tuned using the in-plane gate G3. The QD is coupled to source and drain via two tunnelling barriers, which can be separately controlled with gates G1 and G2. These gates are also used to control the number of electrons in the QD. We use two electrically separated circuits to perform independent conductance measurements through the QPC and the QD at the same time. All measurements are done in a $^3$He/$^4$He dilution refrigerator at a base temperature of 40 mK. In Fig. \ref{fig:qpcprinzip} the conductance of the QD and the QPC is shown as a function of the gate voltage applied to gate G2. The conductivity of the QD (lower line) displays typical Coulomb blockade peaks: Whenever a state in the QD comes into resonance with the leads, a nonzero conductance through the QD occurs. At the same time a step appears in the QPC conductance (upper line) due to the charge of the additional electron on the QD. The steps are superimposed on a gradual rise of the conductance caused by the direct influence of G2 on the QPC potential. Due to its high sensitivity the QPC is an excellent probe for charge redistributions on the QD.

\begin{figure}[tb]
\includegraphics[scale=0.6]{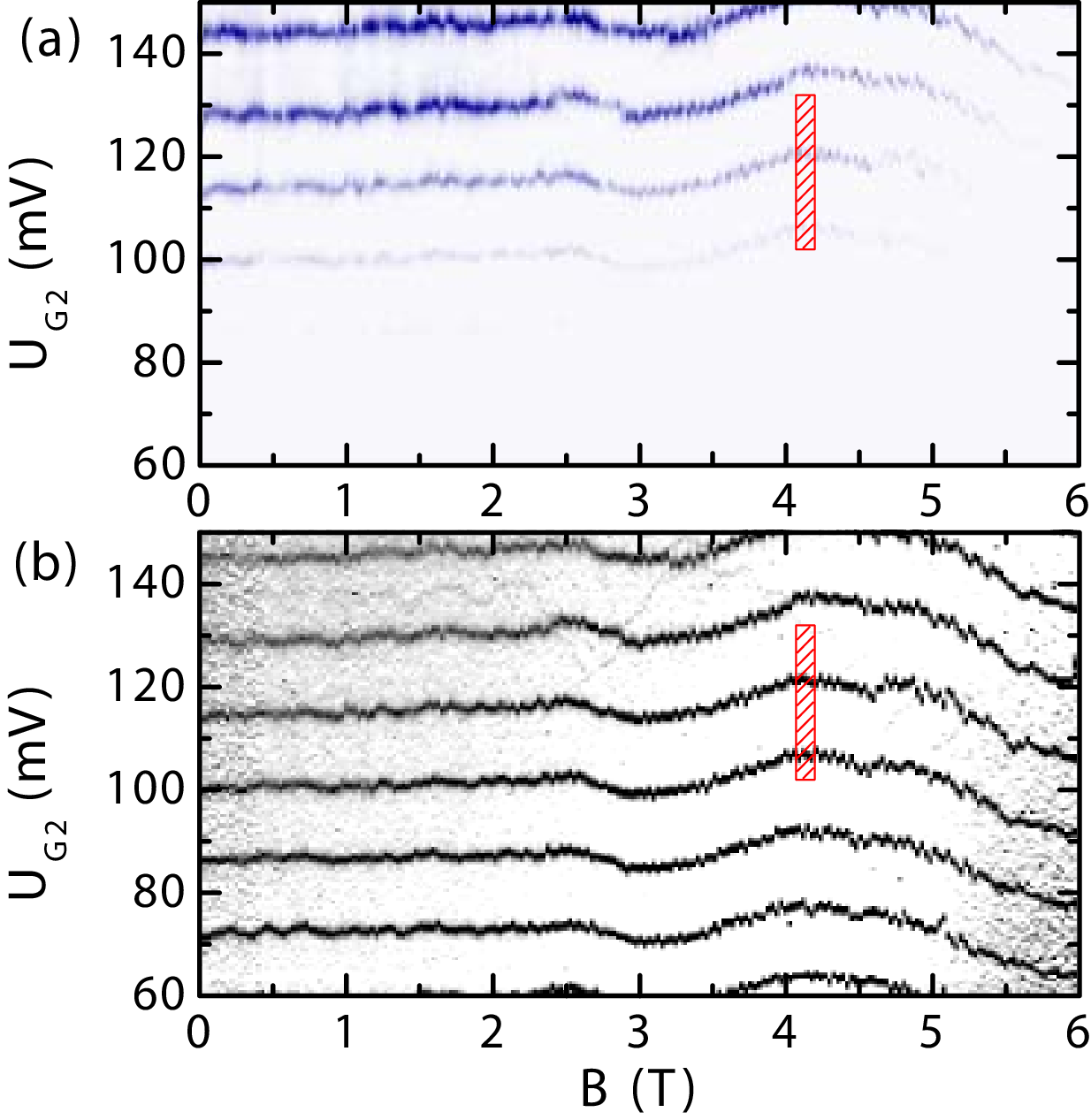}
\caption{\label{fig:uebersicht} (a) Conductance $g\mathrm{_{QD}}$ of the QD and (b) $dg\mathrm{_{QPC}}/dU\mathrm{_{G2}}$ of the QPC as a function of gate voltage and magnetic field, where light-colored means zero conductance and dark finite conductance. Several Coulomb peaks can be seen, changing the position with increasing magnetic field. Even for vanishing transport through the QD the QPC signal clearly traces the addition of further electrons to the QD. The hatched boxes represent the measurement range shown in Fig. \ref{fig:umladungen}.}
\end{figure}

We studied the conductance of the QD and the QPC in dependence of a magnetic field applied perpendicular to the plane of the 2DES. Figure \ref{fig:uebersicht} gives an overview up to 6 T. In Fig. \ref{fig:uebersicht}(a) the conductance of the QD is shown as a function of gate voltage applied to G2 and magnetic field, where white means zero and darker color finite conductance. In the depicted range of gate voltage one can see four darker lines, that represent the Coulomb blockade peaks. The peak positions slightly change with varying magnetic field. This is a result of a change in the Fermi energy of the leads. Between 3 and 5 T the lines show a small zigzag pattern. As there is no transport through the QD between two Coulomb peaks the conductance between the lines is zero. Figure \ref{fig:uebersicht}(b) shows the simultaneously measured QPC signal. We have plotted the derivative of the QPC conductance with respect to the gate voltage, where black means a large change in the conductance and white no change. For each Coulomb peak in the QD conductance, i.e. each change of electron number on the QD, a distinct peak in the derivative of the QPC conductance occurs. The signal shows the same shape as the measurement through the QD. But the QPC detector still works at low gate voltage, when the transport signal $g\mathrm{_{QD}}$ vanishes.

\begin{figure}[tb]
\includegraphics[scale=0.95]{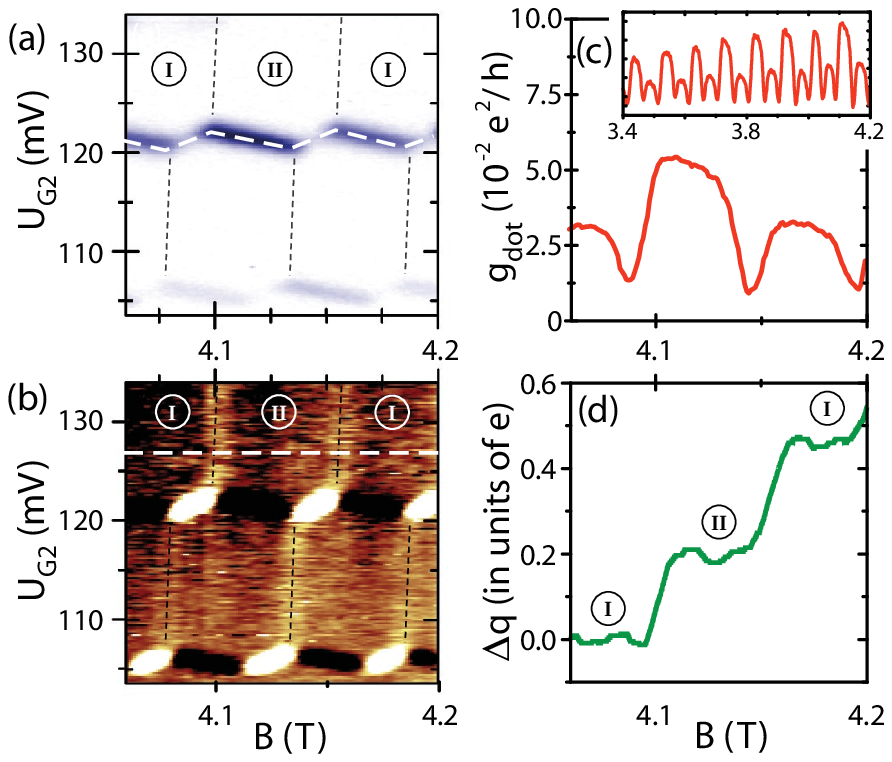}
\caption{\label{fig:umladungen} (a) Conductance of the QD as a function of gate voltage and magnetic field in more detail. The Coulomb peaks show a typical zigzag pattern as a function of magnetic field. (b) Same range of $B$ for $dg_{QPC}/dB$. Vertical dashed lines drawn are guides to the eye. Between the Coulomb peaks vertical lines can be seen, when the electrons rearrange due to the magnetic field. (c) Trace along a Coulomb peak in the QD conductance [dashed line in (a)]. The conductance oscillates over a wide magnetic field range as demonstrated in the inset. (d) $\Delta q$ detected by the QPC as a function of the magnetic field along the dashed line in (b). For each of the light vertical lines a distinct step in $\Delta q$ occurs.}
\end{figure}

In the following we will analyze this behavior in more detail. In Fig. \ref{fig:umladungen}(a)(zoom of hatched box in Fig. \ref{fig:uebersicht}) the typical zigzag caused by the shift of the dot-states in the magnetic field can be clearly observed. 

This behavior is explained as follows. In the shown range of the magnetic field, the spectrum of the QD consists of two Landau levels. The states of the first (the outer) Landau level drop in energy with increasing magnetic field, the states in the second (the inner) Landau level rise. Therefore the energetically favorable state for an electron can be found alternately in the inner Landau level and in the outer Landau level. As a result electrons are redistributed with changing magnetic field and the Coulomb peaks shift upward in gate voltage, whenever the occupied state is in the inner Landau level and shifts downward in gate voltage whenever the occupied state is in the outer Landau level. This leads to the oscillating Coulomb peak position, as seen in \ref{fig:umladungen}(a). 

Moreover, the conductance slightly differs between two neighboring downward line segments. This can be explained by the appearance of spinblockade as discussed by Ciorga et al. \cite{ciorga-00} and recently shown for LAO devices in \cite{max_spin}.

In Fig. \ref{fig:umladungen}(c) a trace along a Coulomb peak [dashed line in Fig. \ref{fig:umladungen}(a)] is shown. It can clearly be seen, that the spin blockade pattern appears: the conductance along the Coulomb peak line shows an oscillating amplitude.

As the electrons on the QD are redistributed with increasing magnetic field, a slight change of the effective charge measured by the QPC should be detectable. Therefore some structure is expected between the Coulomb peaks. In Fig. \ref{fig:umladungen}(b) the QPC signal differentiated with respect to the magnetic field is shown as a function of magnetic field and gate voltage. The graph depicts the same parameter range as Fig. \ref{fig:umladungen}(a). Dark colors represent a decreasing QPC conductance, light colors an increasing conductance. Due to the zigzag pattern the Coulomb peak lines are segmented in black and white parts, as the charge on the QD alternates with rising magnetic field for a fixed applied gate voltage. Between the almost horizontal Coulomb lines additional nearly vertical lines are visible, that are weaker than the Coulomb peaks. They lead from the top of an upward-cusp in a Coulomb peak line to the bottom of the right-hand nearest downward-cusp of the Coulomb peak line above. These lines are the result of the transition of a single electron, that is relocated from the inner Landau level to the outer Landau level. We will refer to these lines as relocation lines in the following.

The QPC signal can be converted into an effective charge detected by the QPC. Therefore it is necessary to calibrate the detector. The easiest way is to use the Coulomb peak where the charge on the QD changes by one electron. In this way the horizontal shift of the QPC signal can be translated into a $\Delta q$, the change of the effective charge. Figure \ref{fig:umladungen}(d) shows $\Delta q$ along a cut through Fig. \ref{fig:umladungen}(b) as indicated by the dashed line. Very distinct steps can be seen with a step height of $0.2$ to $0.3$ effective electron charges. 

\begin{figure}[tb]
\includegraphics[scale=1]{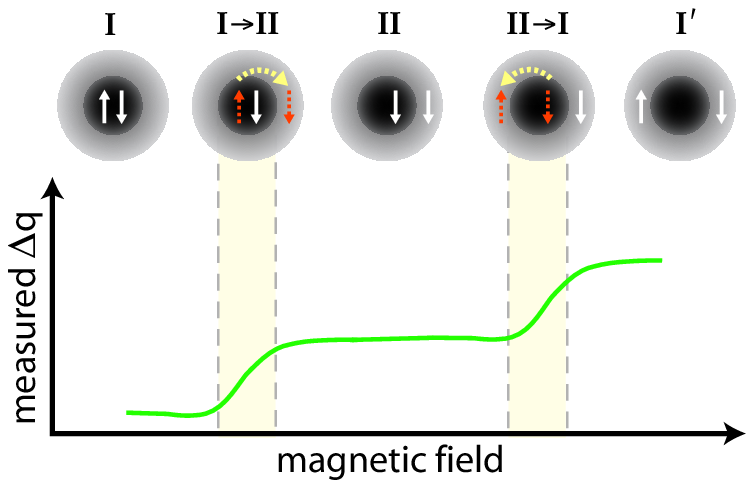}
\caption{\label{fig:model} Simple model for the charge redistribution. The spheres depict the inner and outer Landau shell and the two highest electron states on the QD for the different configurations. Below the idealized QPC signal is shown. The conductance of the QPC slowly decreases due to the raising magnetic field. At each electron transition from the inner to the outer shell a step occurs. Below this the measured charge is shown, as it can be calculated from the QPC signal.}
\end{figure}

A simple picture explaining the charge rearrangement is given in Fig. \ref{fig:model}. We assume that there are two main configurations of the QD called \emph{I} and \emph{II}. The circles show the inner (dark central part of the circles) and outer Landau shell (light ring) of the QD. The electrons occupying the two highest energy states are marked.
In the configuration \emph{I} both electrons occupy the inner shell. With rising magnetic field the energy of the inner shell states increases. At some point it is energetically favorable for the spin up electron to move to the outer Landau shell. This situation is labled \emph{I}$\rightarrow$\emph{II}. When the electron jumps to the outer shell, the potential of the QPC changes slightly. The change in the electron configuration changes the local potential of the QPC. This leads to a significant change in the conductance of the QPC, which can be translated to a change of the effective charge $\Delta q$, although the real charge of the QD stays constant. 
After the transition of the first electron the QD reaches the configuration \emph{II}. The detected charge stays constant until the second electron moves to the outer shell as shown in \emph{II}$\rightarrow$\emph{I}. Again the redistribution of an electron leads to an abrupt change in the QPC signal. Afterwards both electrons are in the outer shell. This state is called \emph{I'} as it is equivalent to the beginning state. 

With this simple model the position of the relocation lines in Fig. \ref{fig:umladungen}b can be understood. In Fig. \ref{fig:umladungen}(a,b,d) the configurations I and II are labeled for exemplification. Between the horizontal Coulomb peak lines no current flows through the QD and the total charge stays constant, but the QPC continues to detect the rearrangement of the electrons on the QD. Transport measurements through the QD are limited to resonances with the leads, but the QPC can be used to analyze the QD structure even when current through the QD is suppressed due to Coulomb blockade.

\begin{figure}[tb]
\includegraphics[scale=0.5]{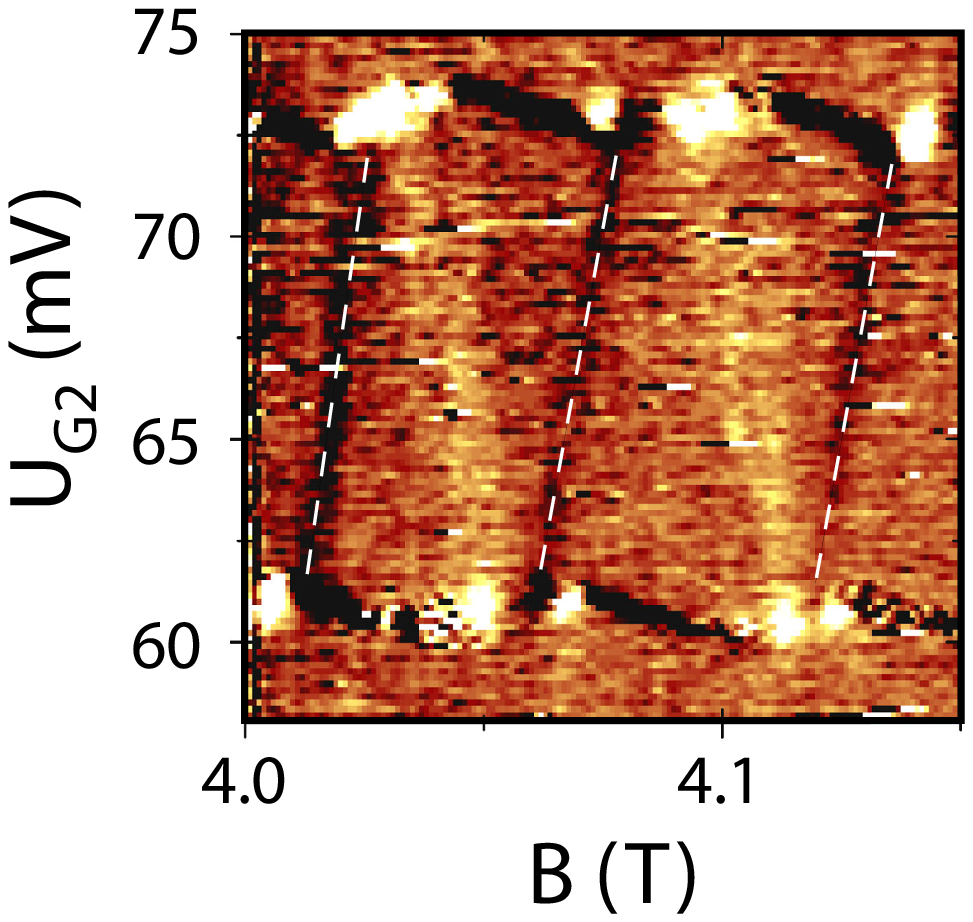}
\caption{\label{fig:umladungen2} $dg\mathrm{_{QPC}}/dB$ as a function of magnetic field and gate voltage. Same type of measurement as presented in Fig. \ref{fig:umladungen}(b) but for a symmetric coupling to source and drain and in a range, where the current though the QD is too low for direct measurement. Again lines can be seen when an electron changes to the outer shell, white dashed lines are drawn as a guide to the eye. Additional bright areas with an inverse slope occur between this lines.}

\end{figure}

For other gate voltages, a rather different behavior can be observed. In Fig. \ref{fig:umladungen2} another $dg\mathrm{_{QPC}}/dB$ plot is shown. In this measurement the conductance of the QD is too low for direct transport measurements. The tunnelling barriers to source and drain are tuned concurrently using the gates G1 and G2 to keep the coupling to source and drain symmetric. Again the zigzag pattern can be seen for the two Coulomb peaks shown. Similar to the measurement presented in Fig. \ref{fig:umladungen}, transition lines are observed. As this form of the QD differs from the configuration with asymmetric barriers shown in Fig. 3b the algebraic sign of the change in charge at the relocation line is different. Here the QPC detects an increasing effective charge for each electron jumping to the outer shell. Slope and position of the lines are similar to the results for asymmetric barriers but additional features can be seen: following each transition line an area of decreasing effective charge appears.  This area shows the opposite slope compared to the relocation lines and leads from an upward flank of the upper Coulomb peak line to an upward flank of the lower Coulomb peak line. The appearance of these areas can not be understood using the simple model presented above. It may result from a selfconsistent rearrangement of the charge inside the QD due to the influence of the magnetic field.

In summary we have performed transport measurements on a quantum dot using a quantum point contact as a charge detector. By this we were able to extend the investigations to a regime, that is unreachable by transport measurements. We analyzed the QD's behavior in high magnetic fields. We used the QPC to observe the QD electron configuration while Coulomb blockade suppresses transport. By this we were able to detect in a non-invasive way the rearrangement of electrons in the QD due to the magnetic field. We analyzed these rearrangement for a constant number of electrons, where only the internal charge configuration changes. We presented a simple model to explain the experimental data and have shown results that need further discussion.


\end{document}